# Understanding Information Hiding in iOS

**Luca Caviglione, National Research Council of Italy**

**Wojciech Mazurczyk, Warsaw University of Technology**

*The Apple operating system (iOS) has so far proved resistant to information-hiding techniques, which help attackers covertly communicate. However, Siri—a native iOS service that controls iPhones and iPads via voice commands—could change this trend.*

Proliferating malware poses significant concerns for victims and security experts alike, and as the number of devices ripe for infection escalates, the amount of at-risk data will only continue to grow. A new form of mobile malware that uses information-hiding techniques to cover up malicious activities has been discovered in Android devices. While iOS devices such as iPads and iPhones have remained relatively secure, we discovered a way to hide information on the iOS system. We present our findings to help security experts better detect potential malware on the iOS platform.

## Background

Information hiding enables attackers to communicate without being noticed by a third-party observer. Originally introduced for written secrets, information hiding is now a key element in effectively assessing network security. Desktop malware that has recently employed information hiding with some success includes Duqu and Alureon[1] (using pictures to transmit stolen data to remote servers), Trojan.Zbot[2] (downloading a jpeg that embeds a list of IP addresses to be inspected), and Linux.Fokirtor[3] (injecting data within Secure Shell traffic to leak information to its command-and-control server).

Because smartphones contain a range of personal and sensitive data, they have become a preferred target for stealing confidential information. Despite the rich set of features that can be exploited via data hiding, only a few malicious applications use these techniques due to mobile appliances' relative newness compared with other computing devices.[4] Recently, though, Android's open source nature allowed hackers to develop a malware called Soundcomber,[5] which covertly transmits the keys pressed during a call (for example, when a user enters his or her PIN to access a financial service).

As iOS increases in popularity, it's also attracting more malware developers. In April 2014,

jailbroken iPhones and iPads were infected via a malicious dynamic library called Unflod.dylib.[6] When running, the library listens to outgoing Secure Socket Layer (SSL) connections to steal a user's Apple ID and password, which are then leaked in plaintext. Although in this instance the absence of stealthiness partially mitigates the threat's effectiveness, it's only a matter of time before we'll see new forms of malware that use information hiding to compromise the Apple ecosystem.

We ourselves discovered a method—which we call iStegSiri—that we believe is the first attempt to covertly leak data from an iPhone or iPad without installing additional applications. Understanding this method can help security experts recognize and prevent similar future attacks.

## Cloaking Information

Information hiding makes it possible to cloak a communication's very existence; thus, it's different from cryptography, in which a transmission's content, though unreadable, is still overt. The two mechanisms are often used jointly—for example, to assure that a conversation remains unreadable. Data hiding is derived from *steganography*, which originally involved techniques like invisible ink or tattoos.[1]

To exchange secrets, the two endpoints must agree on a scheme in advance and embed the secret message within a carrier: the more popular the carrier, the better the masking capacity. Too many alterations would reveal the embedded information's presence, thus limiting the amount of data that can be covertly transmitted. For example, a carrier that injects secrets in the least significant bit (LSB) of a known set of an image's pixels can be discovered due to its visible artifacts.

For attackers exchanging secret data, current network datagrams and sophisticated Internet-scale services offer the ideal choice.[7] While early techniques focused on modifying unused fields of TCP/IP headers (for example, the IPv4 type of service field, which is rarely set by routers), more recent and sophisticated data-hiding methods include exploiting the traffic produced by popular services such as Skype or BitTorrent.[8] From this perspective, modern smartphones offer a variety of new carriers, including cloud services, storage services like Dropbox, and voice-based services like Siri.

## Siri

Originally released as a standalone application in 2010, Siri has been offered as a native iOS service since 2011. It allows users to interact with their iPhone or iPad in two ways: by activating Siri and then giving commands such as creating a note or making a phone call, or by switching at any time from keyboard to voice for entering text. The translation of voice input to text is performed remotely in an Apple-operated server farm. The iPhone or iPad samples the voice, sends it to a remote facility, and waits for a response containing the recognized text, a similarity score, and a time stamp. Figure 1 depicts this usage pattern and architectural blueprint, which lead to an appreciable exchange of traffic between the two parties.[9]

Because Apple has complete control over the application distribution pipeline, the diffusion of information-hiding methods has been efficiently tamed. Still, attackers can use Siri because information hiding doesn't require the device's alteration—or, in fact, any awareness on the device's part—or the installation of additional software components.

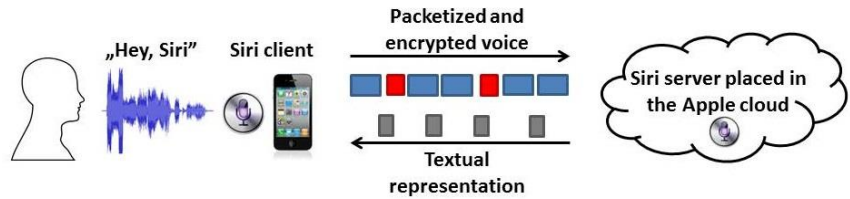

**Figure 1.** Siri's usage pattern and architectural blueprint.

## Method, Limitations, and Countermeasures

Siri processes a user's voice with the Speex Codec, and the related data is transmitted to Apple as a sort of one-way voice-over-IP stream encrypted and encapsulated within HTTP. The main idea behind the method we discovered—iStegSiri—is controlling the "shape" of such traffic to embed secrets. For example, iStegSiri relies solely on specific audio patterns captured by Siri via the hosting device's built-in microphone. Figure 2 depicts a scenario where iStegSiri is used to build a covert channel between an infected device and a botmaster to extract sensitive information (for example, a credit card number or Apple ID and password).[6]

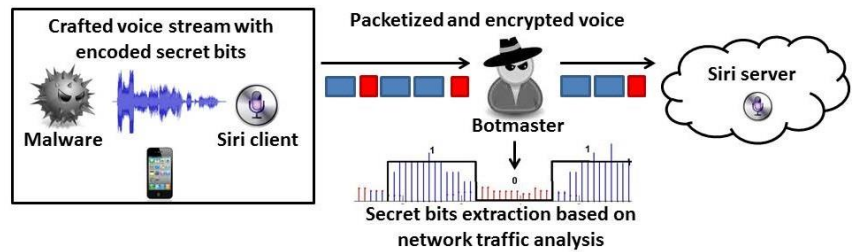

**Figure 2.** Exploiting iStegSiri for malware purposes. Infected smartphone utilizes native Siri service to control the "shape" of the network traffic embedding the secret data.

Specifically, iStegSiri is based on three steps. In step 1, the secret message is converted to an audio sequence based on the proper alternation of voice and silence. In step 2, the sound pattern is provided to Siri as the input via the internal microphone. Consequently, the device will produce traffic toward the remote server that requires audio-to-text conversion. In step 3, the recipient of the secret communication passively inspects the conversation and, by observing a specific set of features, applies a decoding scheme to extract the secret information.

Steps 1 and 2 require properly matching the offered audio and the produced throughput. A set of trials and past measurements[9] demonstrated the feasibility of this approach. Algorithms used for synchronization, latency reduction, and packetization delay prevented forging the shape of the whole flow, even with a minimal degree of accuracy. To overcome this, we split the overall traffic into different components using a set of ranges for Siri's protocol data units (PDUs). PDUs in the range of 800–900 bytes represented talk periods, while PDUs in the range of 100–700 bytes represented inactive periods.

With this partition, we were able to arbitrarily encode 1 and 0 within the traffic. In other words, alternating talk and silence periods increased or decreased the number of PDUs belonging to each defined range. Nevertheless, some form of voice activity detection (VAD) impedes high symbol rates, or the speed at which voice and silence alternates. In our trials, the shortest working values were one second and two seconds for voice and silence, respectively.

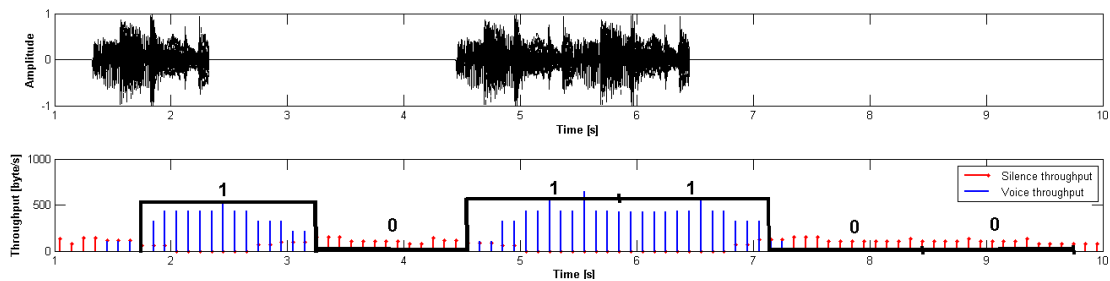

**Figure 3.** iStegSiri's crafted voice stream (a) results in the corresponding classes of traffic (blue=voice, red=silence) (b) which allows to successfully detect secret data bits at the receiving side (b).

To complete Step 3, the covert listener must capture the traffic and decode the secret. The former can be achieved in several ways, including transparent proxies or probes that dump traffic for offline processing. The decoding algorithm implements a voting-like method using two decision windows to determine whether a run of throughput values belongs to voice or silence (1 or 0).

Figure 3 depicts the outcome of a covert transmission. We found that the iStegSiri method can send secrets at a rate of about 0.5 bytes per second. For instance, a typical 16-digit credit card number can be transmitted in about 2 minutes.

An attacker looking to exploit iStegSiri can access Siri functionalities in jailbroken devices through a library called Libactivator, or by directly accessing the private APIs provided by Apple in a plain environment. The malware can produce the audio track used to encode the secret at runtime, for example, by replicating a single sample via software, without having to inflate the size of the executable. Nevertheless, even if we used the microphone for the performance evaluation, audio data can be directly routed from the malware to the codec; it doesn't have to be played back audibly by the user.

As designed, iStegSiri has two main limitations, neither insurmountable for hackers:

- It requires access to Siri's inner workings; this means that only jailbroken iOS devices can currently be used. However, iStegSiri showcases the principle of using real-time voice traffic to embed data. Therefore, it can be further exploited on existing similar applications such as Google Voice or Shazam, or implemented in future applications by taking advantage of coding errors.

- It requires access to the steganographically modified Siri traffic as the traffic travels to server facilities. However, as already suggested, this can be achieved in several ways, including transparent proxies or probes that dump traffic for offline processing. This somewhat shifts the threat from the application to the network, thus proper additional countermeasures must be evaluated.

**B**ecause information-hiding methods use very specific technological traits, no current off-the-shelf products effectively detect covert communications. This forces security experts to craft dedicated countermeasures for each method. With iStegSiri, the ideal countermeasure acts on the server side. For example, Apple should analyze patterns within the recognized text to determine if the sequence of words deviates significantly from the used language's typical behaviors. Accordingly, the connection could be dropped to limit the covert communication's data rate. This approach wouldn't rely on the device, so additional functionalities or battery consumptions wouldn't be required. We plan to further our research to develop an efficient countermeasure to mitigate this threat.

Jan. 2010, pp. 40–45.

**Luca Caviglione** *is a researcher at the National Research Council of Italy. He holds several patents and is an associate editor of Wiley's* Transactions on Emerging Communications Technologies. *Caviglione received a PhD in computer science from the University of Genova.*

**Wojciech Mazurczyk** *is an associate professor at the Institute of Telecommunications in the Department of Electronics and Information Technology at the Warsaw University of Technology. He is also an associate technical editor for* IEEE Communications Magazine. *Mazurczyk received a PhD and DSc in telecommunications from the Warsaw University of Technology.*